# Perspective: Plasmon antennas for nanoscale chiral chemistry

Esteban Pedrueza-Villalmanzo [1], Francesco Pineider [2] and Alexandre Dmitriev [1,*]

[1] *Department of Physics, University of Gothenburg, Göteborg, Sweden*

[2] *Department of Chemistry and Industrial Chemistry & INSTM, University of Pisa, Pisa, Italy*

* *alexd@physics.gu.se*

Abstract

Plasmon nanoantennas are extensively used with molecular systems for the chemical and biological ultra-sensing, for boosting the molecular emissive and energy transfer properties, for the nanoscale catalysis, and for building advanced hybrid nanoarchitectures. In this Perspective we focus on the latest developments of using plasmon nanoantennas for the nanoscale chiral chemistry and for advancing the molecular magnetism. We overview the decisive role nanoplasmonics and nano-optics can play in achieving the chirally-selective molecular synthesis and separation, and in the way such processes might be precisely controlled by potentially merging chirality and magnetism at the molecular level. We give our view on how these insights might lead to the emergence of exciting new fundamental concepts in the nanoscale materials science.

Chirality, or structural handedness, is a fundamental feature of biological life. It refers to the chemically identical molecular species having non-superimposable structural arrangement. It is then said such molecular species have one or another (R or S) handedness. This lends an extremely high specificity to a host of various molecular interactions and to a build-up of complex molecular architectures that are the fundamental building blocks in life science, biomedicine and pharmacological industry. Asymmetric chemical synthesis caters to these extensive needs and is employed to produce the molecular species with designed handedness, that is, an enantiomeric excess in the final product of a chemical reaction. Various catalysts are used to promote the emergence of the enantiomeric excess, typically in the same (liquid) phase as the reactants (homogeneous asymmetric catalysis). This usually complicates the task of separating the products from the catalyst after the reaction. Heterogenous (surface-based) catalysis then comes in help [1]. A convenient way of measuring the enantiomeric excess and, in general, the molecular chirality, is by detecting the preferential absorption of right- or left-circularly polarized light (RCP or LCP) that passes the sample. Such preferential absorption is expressed as the circular dichroism (CD): $CD \propto A_{RCP} - A_{LCP}$. Another parameter that is often employed to characterize the molecular chirality is the Kuhn dissymmetry factor $g$, also expressed through the light circular polarization-dependent absorption as $g = \frac{2(A_{RCP} - A_{LCP})}{A_{RCP} + A_{LCP}}$. However, solutions with low enantiomeric excess naturally suffer from the low $g$ and, in principal, low CD at low molecular concentrations. As a consequence, extensive research efforts are directed both at generally increasing the enantiomeric excess of the products of the asymmetric synthesis and at substantially boosting the detection sensitivity of the chiral molecular species. The latter is widely featured in the recent literature [2-4], and in this Perspective we focus on the former, that is, the recent advances and prospects of driving the chiral chemistry at the nanoscale aimed at achieving sizeable enantiomeric excess. Figure 1 summarizes the discussion that follows: we overview four various paths of producing the enantiomeric excess with nano-confined asymmetric chemical reactions, namely employing superchiral electromagnetic fields in plasmon cavities; Fano-type resonances; propagating surface plasmon polaritons; and the chiral induced spin selectivity.

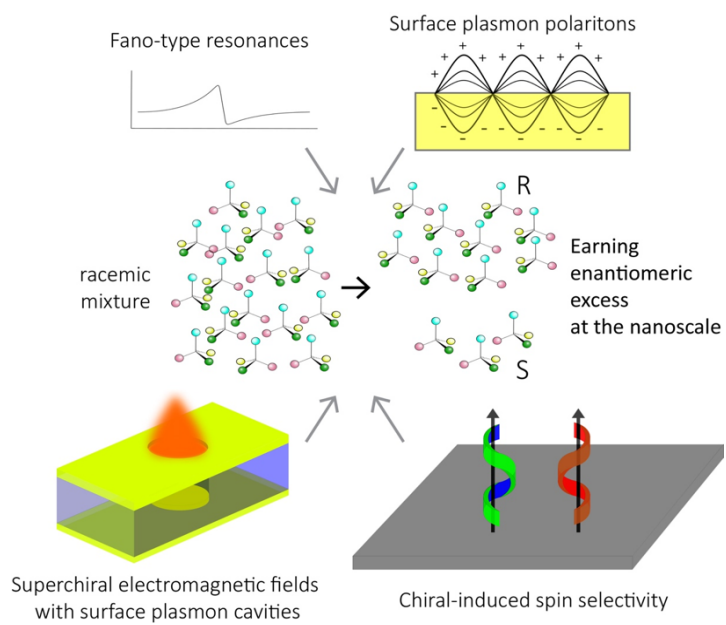

*Figure 1. Nanoscale chiral chemistry: producing the enantiomeric excess with nano-confined asymmetric chemical reactions by the superchiral electromagnetic fields in plasmon cavities; Fano-type resonances; propagating surface plasmon polaritons; and the chiral induced spin selectivity.*

The CD spectroscopy is a staple spectroscopic technique to detect and quantify enantiomeric excess. Interestingly, it has been demonstrated that the use of circularly polarized (CP) light of either handedness in photochemical reactions can in itself promote the formation of one enantiomer over the other simply by the selective absorption of light with different handedness [5, 6]. Unfortunately though, as it is pointed out in [7], the actual chiral selectivity and the yield of such reactions is rather low, around 2%. In Fig. 2b, it is possible to appreciate the small signature on the CD measurement (less than 1 mdeg) that represents an enantiomeric excess of 0.4 % [8]. One interesting example of the generation of CP light-induced chirality, with addition of a second irradiation by CP UV ultra-violet light (CPUL) to lock the obtained configuration is showed in Fig. 2a [9].

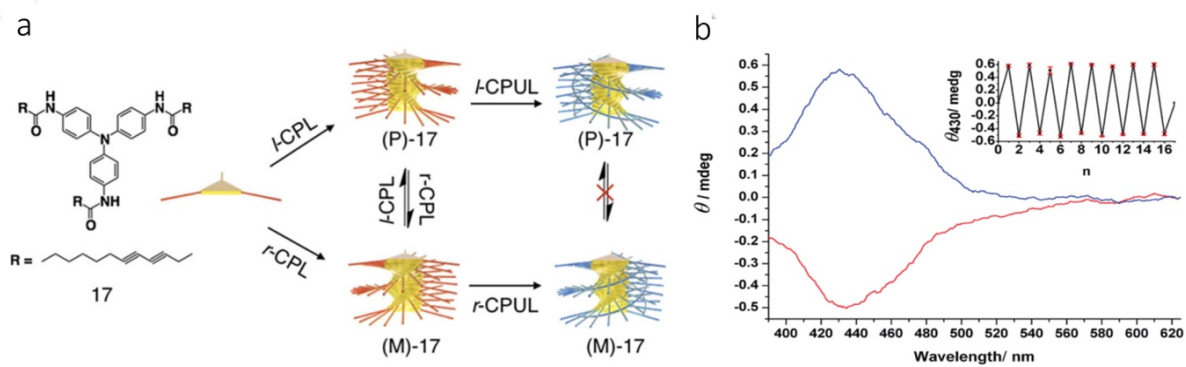

*Figure 2.* (a) Induction, control and locking of supramolecular chirality by CPL and CPUL from an achiral compound. Reproduced from Ref. [10]; (b) Induced CD spectra of a photoswitch molecule upon irradiation with r-CP light (blue) and l-CP light (red) at 436 nm. Reproduced from Ref. [8].

The recent advances in the asymmetric (photo)synthesis and chiral amplification, that is, total photo-conversion of a racemic mixture into one pure enantiomer, are reviewed in [10]. One of the implementation are the CP light-triggered asymmetric autocatalysis reactions, showing to be of great potential in achieving nearly enantiomerically pure solutions [11-13]. Another modality of separating the enantiomers is using optofluidics. Tkachenko et al. [14] developed a system of sorting chiral micrometer-size droplets depending on their handedness and using counterpropagating beams with RCP and LCP light. This approach could be used to sort smaller objects as well by using the droplets as "conveyors" and employing appropriate surface functionalization discriminating between chiral and non-chiral objects (molecules or particles). A promising approach is the use of (metal) nanoparticles to separate enantiomers in solution. While several methods rely on the use of Au [15, 16] and Ag nanoparticles [17-19], these do not engage the plasmon resonances, but rather the functionalization of the metallic surface with chiral or non-chiral ligands [15, 17], also exploring the use of the inherent chirality of Au or Ag nanoparticles or clusters [18, 16, 19] (note a helpful review in [20] on the subject).

Considering the encouraging results demonstrated by the CP light-driven emergence of the enantiomeric excess in asymmetric synthesis, it's only natural to take the advantage of the intense light-matter interaction at the nanoscale afforded by the plasmon resonances of metallic nanostructures in solutions or in metasurfaces. Molecular species can be found in very close contact to such nanostructures without forming a chemical bond and taking advantage of the highly enhanced electromagnetic (super)chiral near-fields in the direct proximity to plasmon chiral or achiral resonators (nanoantennas). Several works have approached this exciting prospect from a theoretical point of view, yet no experimental evidence of plasmon-enhanced photochemical asymmetric synthesis has been reported so far.

Superchiral electromagnetic near-fields (that is, the electromagnetic fields with the optical chirality higher than that of a free-propagating CP light) can be induced in the direct proximity of the nanostructures, interacting with the CP or even linearly-polarized light. Such nanostructures may form extended arrays as photonic (semiconductor), plasmonic (nanometallic), or dielectric metasurfaces. The striking characteristic of these metasurfaces is that they often feature achiral motif. The chiral electromagnetic near field links to the optical chirality $C$ as [21, 22] : $C\{\boldsymbol{E},\boldsymbol{H}\} = \frac{-k_0}{2c_0} Im\{\boldsymbol{E} \cdot \boldsymbol{H}^*\}$ where $\boldsymbol{E}$ and $\boldsymbol{H}$ are the electric and magnetic fields of light, and $k_0$ and $c_0$ are the wavevector and the speed of light in free space, respectively. The principle idea behind the electromagnetic design of a nanoantenna aimed at efficient generation of the enantiomeric excess during the chiral photo-transformation at the nanoscale is to maximize $C$. To achieve this, certain conditions need to match like the spectral and spatial overlap of the electric and magnetic near-fields while having $\pi/2$ phase difference between them [21-23]. Several examples of metasurfaces realize these conditions using plasmon nanoantennas. Vázquez-Guardado et al. [24] provided a nanohole-nanodisk metasurface supporting two degenerate localized surface plasmon electric and magnetic modes. Here the electromagnetic cavity mode, formed by the nanodisk at the bottom of the nanohole and the metallic nanohole rims, deliver a strongly chiral near-field that is designed to spectrally overlap with a C-H vibrational mode of a chiral molecule (camphor), experimentally improving its detection sensitivity by 4 orders of magnitude. Rui et al. [25] theoretically proposed a metal-dielectric-metal metasurface with the 1300-fold boost of the molecular CD coming from the gap in the nanostructures. Other works have proposed Ag (Fig. 3a) and Au (Fig. 3d) plasmonic nanoapertures for the same purpose [26, 23].

Using the high-refractive-index semiconductor and dielectric nanoantennas presents the advantage of having strong electric and magnetic electromagnetic dipoles within the same nanostructure with the added benefit of low losses and high resonator quality factors. Such nanoantennas have been proposed for the chiral near-fields generation by J. Dionne et al. [7] (Fig. 3c). An intriguing property of such high-index nanoantennas is the occurrence of the so-called generalized Kerker condition, where higher order modes interfere constructively, specifically, the in-phase electric and magnetic modes are spatially and spectrally overlapping, to produce highly directive forward light scattering [27, 28]. This effect has been proposed for the enantiomeric excess generation by the near-field of the semiconductor nanostructures [22, 29]. Calculations made by Solomon et al. [29] unearthed that a metasurface made of GaP nanodisks could potentially have photolysis reaction yields for a thiocamphor up to 30% with an enantiomeric excess of 10%, as compared to the asymmetric reaction yield of 0.66% using just CP light. This is achieved by the 4.2-fold increase of the intrinsic molecular dissymmetry factor when the molecules are above the surface of a nanodisk (Fig. 3b).

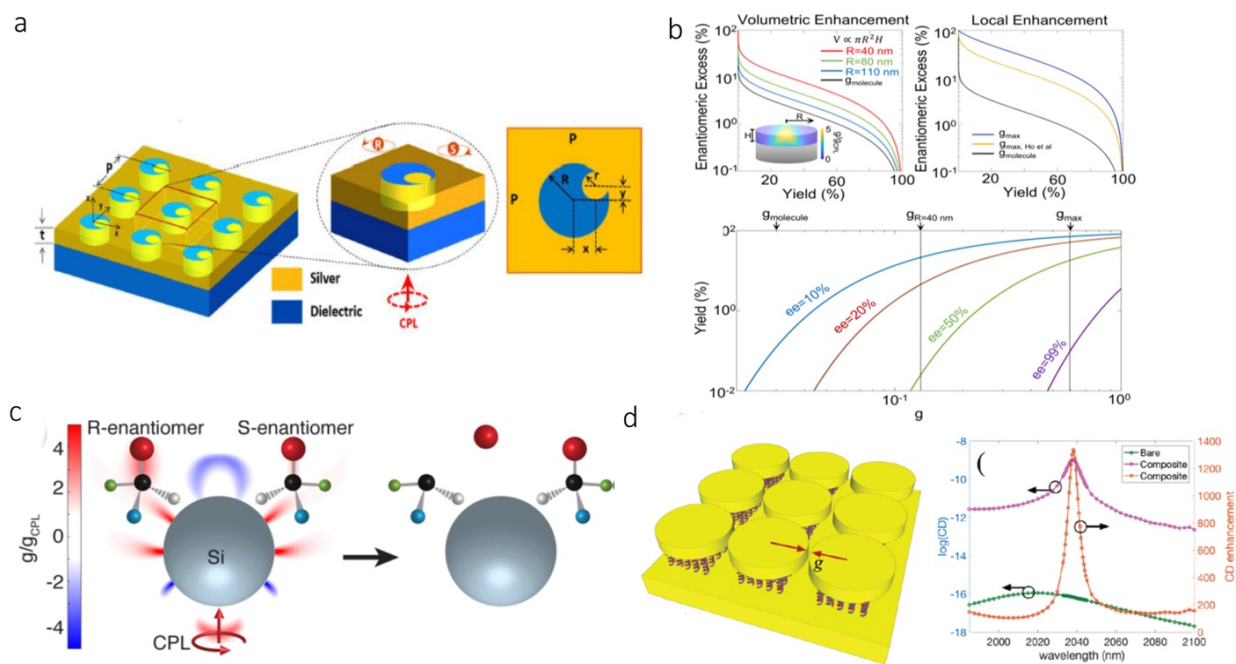

*Figure 3.* (a) Schematic representation of the plasmonic device formed by a periodic array of asymmetric nanoapertures for enantioselective optical process. Reproduced from [26]. (b) (upper left panel) Enantiomeric excess plotted against percent yield in a photoionization reaction based on thiocamphor (g = 0.04) for three different volume regions above a disk with fixed height (H =40 nm); (upper right panel) Enantiomeric excess plotted against percent yield in a photoionization reaction based on thiocamphor for no enhancement, the maximum point enhancement for a silicon sphere, and that for the GaP disk metasurface (7- and 15-fold, respectively); (lower panel) Percent yield plotted against absolute g for enantiomeric excesses (ee) of 10%, 20%, and 50%. Reproduced from [29]. (c) Schematics for photolysis of a molecule near a silicon nanosphere of radius r illuminated by circularly polarized light. Enhanced preferential absorption near the nanosphere excites a vibrational mode in the right-handed (R) enantiomer, leading to dissociation of one bond while leaving the left-handed (S) enantiomer intact. Reproduced from [7]. (d) (left) Conceptual illustration of chiral molecule/MDM coupled system; (right) The CD and CD enhancement spectra for a chiral molecule layer with and without the nanogap metamaterial absorber structure. Reproduced from [25].

Fano-type resonances that are typically resulting from the interacting of the high-quality (optical) resonator and a spectrally broader background (alternatively – lower order resonances) have been proposed for enantioseparation by using the electromagnetic near-field. Specifically, using dipole-octupole Fano-type resonances a transverse optical force appears and optomechanical chiral sorting by trapping of sub-10 nm chiral particles (Fig. 4a) becomes available [30, 31]. Also focusing on the optical forces, the use of propagating surface plasmons has been reported [32] (Fig. 4b). Several conceptual works utilized quantum spin Hall effect [33, 34] employing the evanescent optical fields, where chiral particles with different handedness experience opposite lateral forces, which can also be realized at a metal-dielectric interface [35]. Similarly, optical tweezers are used as a very accurate enantioselection tool for the chiral nanoparticles, where optical trapping is dependent on the nanoparticles structural chirality. Zhao et al. [36] recently demonstrated that plasmonic optical enantioselective trapping with a metallic tip is achieved at 20 nm distance from the tip. The same research group (J. Dionne) introduced a technique to potentially separate enantiomers using a chiral atomic force microscope probe coupled to a plasmonic optical tweezer that exerts the attractive or repulsive lateral forces, depending on the handedness of the employed CP light, with the differences in forces up to 10 pN [37].

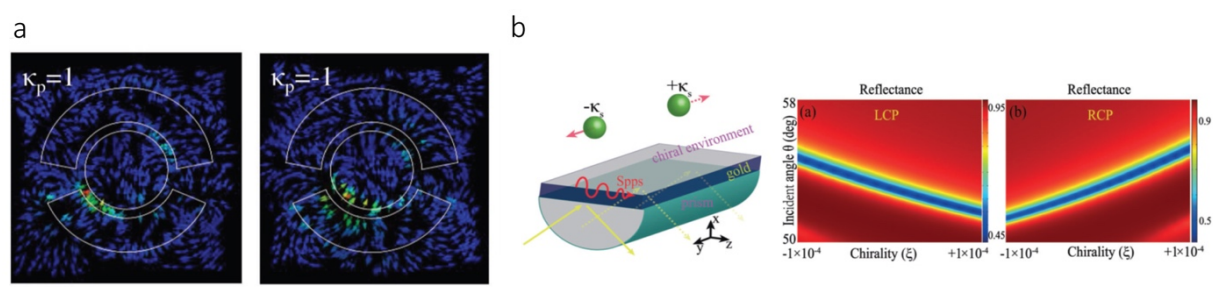

*Figure 4. (a) The total transverse force on the paired enantiomers with right-hand (left) and left-hand (right) chirality from a complex Fano-nanoantenna. Note how the chiral molecules with different helicities are separated for the gap. Reproduced from [30]. B) (left) Schematic representation of the metal–chiral Kretschmann configuration for SPP-assisted chiral enantiomer discrimination and separation. (right) Maps of reflectance as a function of the incident angle of LCP or RCP waves and the chirality of the dielectric medium . Reproduced from [32].*

It is highly intriguing to further explore the coupling of the molecular resonances to the electromagnetic superchiral cavities, mentioned above as the means of achieving a substantial enhancement of the chiral molecular species detection.  We may ask ourselves if further increasing the coupling strength in such system can favor the actual enantioseparation or even promote the enantiomeric excess of the optical near-field-controlled asymmetric synthesis. A seminal paper by Hutchison et al. [38] demonstrated the sizeable modification (specifically, slowing down) of the chemical reaction rate of a phototransformation when run under conditions of the so-called 'strong coupling' of the molecular excitons and the optical mode of a Fabry-Perot cavity. Further perspective on the chemistry involving strong coupling (so called 'polaritonic chemistry') is given in [39]. Indeed, as already pointed out in the work of Vázquez-Guardado et al [24], the dissymmetry factor can increase when the cavity is tuned to the vibrational band of a chiral molecule. Although this is a surface-based enhancement phenomena, the principle of using superchirality (like it is described in [7]) to rate-selectively photo-decompose or in other way photo-transform one enantiomer in the mix may become available. Here the strong coupling of the vibrational and/or optical molecular modes to the vibrational and/or optical superchiral modes of the meticulously designed electromagnetic cavity or nanoantenna would be essential.

It is also stimulating to explore the potential connection of the chiral selectivity to another phenomenon – that of magnetism. The latter is intimately connected to the molecular chiral phenomena. In 2018, the Naaman group [40] demonstrated a selective absorption of the chiral molecules in a racemic mixture using magnetized achiral substrates, proposed to be driven by the chirality-induced spin selectivity. An interesting first step of merging the magnetism at the molecular scale and optical nanoantennas is proposed in the recent work by Pineider et al [41]. It demonstrates an enhancement in the magneto-optical activity of $TbPc_2$, a model single-molecule magnet (SMM), when combined with a simple plasmon antennas (Au nanodisks) (Fig. 5a). It has been demonstrated that plasmon antennas can be used conveniently for boosting magneto-optics in a variety of configurations [42-45]. In the aforementioned work, the plasmon antenna resonance is spectrally overlapping with the excitonic absorption of the $TbPc_2$, which results in a 5-fold plasmon-enhanced molecular absorption (measured by magnetic circular dichroism, MCD) and similarly enhanced magneto-optical activity of the SMM (Fig. 5b and 5c). Above we outlined the scenario of the strong molecular exciton-plasmon cavity coupling. Here, however, this is not achieved and the system studied by Pineider et al. can be treated as a linear superposition of the molecular and the plasmonic oscillators. The reason behind this behavior could be that an additional requirement for the onset of the strong coupling phenomena, besides the energy match, the spatial superposition and the requirements on the electromagnetically non-leaky character of the cavity and a high-quality (low loss) of the resonator (in general – a two-level system) in the cavity, is that the mode volume of the both a cavity and a resonator should be comparable [46].

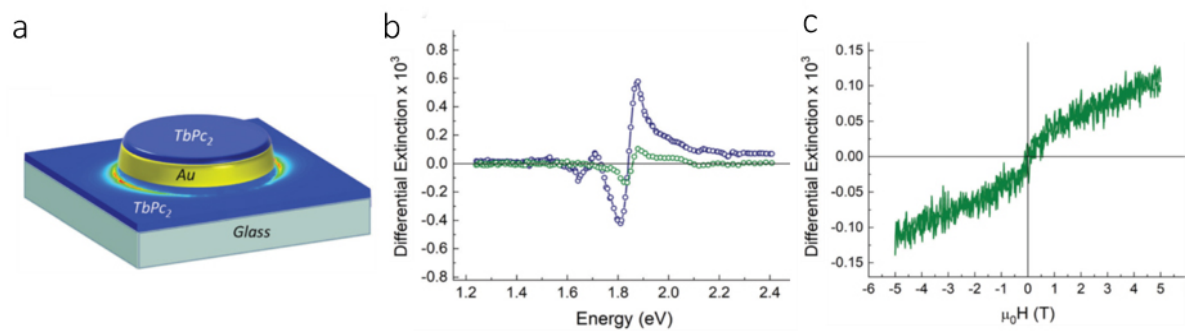

*Figure 5 (a) Schematic representation of the TbPc2 layer on gold nanoantennas overlaid with finite difference time-domain (FDTD) simulations of the electromagnetic nearfield around the nanoantenna. (b) Molecular part of the MCD spectrum of TbPc2@Au corrected for the nanoantenna contribution*

*(blue) and compared to the control MCD spectrum of TbPc2 (green). (c) Isolated TbPc2 hysteresis loop acquired at the positive maximum of the Q band (E = 1.87 eV) when deposited on top of the Au discs. Reproduced from [41].*

In the case of a simple nanodisk antenna, the mode volume could be approximated to the nanostructure itself (several thousands of nm$^3$). The mode of volume of the molecular resonance, on the other hand, depends on the degree of delocalization of the molecular exciton; while no in-depth study has been presented on this matter, one can reasonably assume that delocalization is small or absent, considering that the resonance of the TbPc$_2$ system suffers no significant spectral shift compared to that of isolated molecules in solution. It is then reasonable to assume that the mode volume of the molecular resonance roughly coincides with that of the molecule (that is, in sub-nm$^3$ regime). It is then apparent that the mode volumes are not matching well in the TbPc$_2$-nanodisk antenna system. In this framework, the MCD signal resulting from the hybrid system consists of a convolution of the magneto-optical signal from TbPc$_2$ and a relatively strong contribution from the plasmon antenna. The latter, despite being diamagnetic, exhibits a sizeable magneto-optical activity arising from the perturbation induced by the applied magnetic field on the oscillation of charge carriers involved in the plasmon resonance [47, 48]. Since this effect is qualitatively and quantitatively well-understood, it becomes possible to model and subtract the plasmon contribution to the MCD signal, thus obtaining the plasmon-enhanced magneto-optical signature of SMMs. Interestingly, using the different magnetic behavior of the molecules (saturating their magnetization with increasing the applied magnetic field) and of the nanoantennas (linear to the applied magnetic field), it is possible to discriminate the nature of the MCD signal in different portions of the spectrum. Recently, the group of Liz-Marzán reported magneto-optical experiments on an Au nanorod antenna-dye assembly exhibiting a strong coupling phenomenon [49]. The authors found that the (non-magnetic) dye molecule, characterized by the negligible magneto-optical activity in its native form, starts displaying a sizeable MCD spectral contribution when adsorbed on the Au nanorod antennas. This effect has been ascribed to the partial hybridization of the molecular optical resonance to that of the nanorod (which is MCD active, as described above). This example demonstrates a convincing strategy of transferring the properties while in the strong coupling regime, which might potentially include the electromagnetically chiral interactions.

Looking forward, it is exciting to anticipate how exploring the various molecule-cavity coupling regimes, including the strong coupling with its marked emergence of the hybrid molecular states allowing to bypass the expected chemical reaction routes, and the intriguing connection between molecular chirality and magnetism, might eventually contribute to the development of not foreseen avenues towards the highly efficient chiral chemistry at the nanoscale.

**Acknowledgements**

AD and EPV acknowledge the Knut and Alice Wallenberg Foundation for the project "Harnessing light and spins through plasmons at the nanoscale" (2015.0060). This work has furthermore been supported by the European Union's Horizon2020 Research and Innovation program, Grant agreement No. 737709 (FEMTOTERABYTE). EPV acknowledges the Swedish Energy Agency (Energimyndigheten), project P42028-1.